\documentclass{PoS}

\usepackage{amsmath,amssymb,wrapfig}

\newcommand{\Bs}{{\text{B}_\text{s}}}
\newcommand{\K}{\text{K}}
\newcommand{\C}{\mathcal{C}}
\newcommand{\tk}{{t_\K}}
\newcommand{\tb}{{t_\Bs}}

\newcommand{\R}{\mathcal{R}}
\newcommand{\Ri}{\mathcal{R}^{\,\textrm{I}}}
\newcommand{\Rii}{\mathcal{R}^{\,\textrm{II}}}
\newcommand{\Riii}{\mathcal{R}^{\,\textrm{III}}}

\newcommand{\CBK}{\C^{\Bs\to\K}}

\newcommand{\eq}[1]{Eq.~(\ref{#1})}

\newcommand{\EK}[1]{E_{\rm K}^{(#1)}}
\newcommand{\EB}[1]{E_{\rm \Bs}^{(#1)}}

\newcommand{\tfit}[2]{t^{\rm #1}_{\rm #2}}  

\title{Extraction of the bare form factors for the semi-leptonic $\Bs$
decays}

\ShortTitle{Extraction of the bare form factors for the semi-leptonic $\Bs$
decays}

\author{F.~Bahr, D.~Banerjee, \speaker{M.~Koren},
H.~Simma, R.~Sommer\\
NIC, DESY, Platanenallee 6, D-15738 Zeuthen, Germany\\
E-mail: \email{felixtbahr@googlemail.com},
\email{debasish.banerjee@desy.de},
\email{mateusz.koren@desy.de}, \email{hubert.simma@desy.de}, 
\email{rainer.sommer@desy.de}}

\abstract{The computation of the form factors for the $\Bs\to\K\ell\nu$ decay is
presented. The b quark is treated by means of Heavy Quark Effective Theory,
currently in the static approximation. In these proceedings we discuss the
extraction of the bare matrix elements from lattice data through a combined fit
to two- and three-point correlation functions, as well as by considering
suitable ratios. The different methods agree concerning the extracted form
factors and approximately 2\% accuracy is reached. The non-perturbative
renormalization and matching to QCD is described in accompanying proceedings
\cite{deb16}.}

\FullConference{\mbox{\hspace{12.3cm} \textnormal{DESY 16-251}}\\
    34th annual International Symposium on Lattice Field Theory\\
    24-30 July 2016\\
    University of Southampton, UK}

\begin{document}

\section{Introduction: definitions, ensembles, measurements}

The QCD matrix elements for the semi-leptonic $\Bs\to\K\ell\nu$ decay in the
rest frame of the $\Bs$ meson are
\vspace{-0.2cm}
\begin{align}
(2m_\Bs)^{-1/2}\langle\K(\vec{p}_\K)|V^0(0)|\Bs(0)\rangle &=
h_\parallel(E_\K),\\
(2m_\Bs)^{-1/2}\langle\K(\vec{p}_\K)|V^k(0)|\Bs(0)\rangle &= p^k_\K
h_\perp(E_\K),
\end{align}
where $V^\mu(x) = \bar\psi_{\rm u}(x)\gamma^\mu\psi_{\rm
b}(x)$ and $\vec{p}_\K$ is the Kaon momentum.

The b quark is treated in the framework of Heavy Quark Effective Theory (HQET)
where a full non-perturbative renormalization program exists \cite{strategy}. Here
we focus on the extraction of the bare HQET matrix elements from the lattice
data, while the procedure to obtain the QCD form factors from these bare matrix
elements is described in \cite{deb16}.

For the heavy quark we use HYP1 and HYP2 discretizations \cite{hyp}. We first
restrict ourselves to the static approximation. The two- and three-point
functions of interest are
\begin{align}
  \C^\K(\tk;\vec{p}_\K) =& \sum_{t_i}\sum_{\vec{x}_f,\vec{x}_i}
  e^{-i\vec{p}_\K\cdot(\vec{x}_f-\vec{x}_i)}\langle
  P_{\rm su}(\vec{x}_f,t_i+t_\K)P_{\rm us}(\vec{x}_i,t_i)\rangle,\\
  \C^\Bs(\tb;\vec{0}) =& \sum_{t_i}\sum_{\vec{x}_f,\vec{x}_i}\langle
  P_{\rm sb}(\vec{x}_f,t_i+t_\Bs)P_{\rm bs}(\vec{x}_i,t_i)\rangle,\\
\begin{split}
  \C^{\Bs\to\K}_\mu(\tk,\tb;\vec{p}_\K) =&
  \sum_{t_i}\sum_{\vec{x}_f,\vec{x}_v,\vec{x}_i}
  e^{-i\vec{p}_\K\cdot(\vec{x}_f-\vec{x}_v)} \langle P_{\rm
  su}(\vec{x}_f,t_i+t_\Bs\!+t_\K) V_\mu(\vec{x}_v,t_i+t_\Bs) P_{\rm
  bs}(\vec{x}_i,t_i) \rangle,
\end{split}
\end{align}
with $P_{\rm q_1q_2}(\vec{x},t) = \bar\psi_{\rm q_1}(\vec{x},t)\gamma_5
\psi_{\rm q_2}(\vec{x},t)$. For the light quarks we use Wuppertal smearing
\cite{gauss,mb13}. The Kaon correlator is calculated with only one level of
smearing, while for the $\Bs$ meson we apply three levels of smearing (in both
the two- and three-point functions). We can decompose the Euclidean correlation
functions as
\begin{align}
  \C^\K(\tk) &= \sum_m(\kappa^{(m)})^2e^{-E_\K^{(m)}t_\K},\label{eq:c2ll}\\
  \C^\Bs_{ij}(\tb) &= \sum_{n}\beta_i^{(n)}\beta_j^{(n)}e^{-E_\Bs^{(n)} t_\Bs},
  \label{eq:c2hl}\\
  \C^{\Bs\to\K}_{\mu,i}(\tk,\tb)
  &=\sum_{n,m}\kappa^{(m)}\varphi^{(m,n)}_\mu\beta_i^{(n)} e^{-E_\K^{(m)} t_\K}
  e^{-E_\Bs^{(n)} t_\Bs},\label{eq:c3}
\end{align}
where the indices $m,n$ label the Kaon and $\Bs$ meson energy levels
respectively, while the indices $i,j$ label the smearing levels used for the
$\Bs$ meson. In our current setup $\vec{p}_K$ has a non-zero component only in
the $x$-direction, therefore we only extract the form factors for $\mu=0,1$.
Thus, the desired static bare matrix elements are given by
\begin{equation}
h^{\rm stat,bare}_\parallel = \varphi^{(0,0)}_0\sqrt{2E^{(0)}_\K}, \quad
p^1_\K h^{\rm stat,bare}_\perp = \varphi^{(0,0)}_1\sqrt{2E^{(0)}_\K}.
\end{equation}

We use three $N_f=2$ CLS ensembles \cite{cls2}: A5, F6, and N6, which have
similar pion mass ($m_\pi=$310--340 MeV) but different lattice spacings
($a\approx0.075$, 0.065 and $0.048\;\text{fm}$ respectively), allowing us to
take the continuum limit. For further details on the ensembles, see Table 2 of
Ref.~\cite{paper2}.

We choose $|\vec{p}_\K| = 0.535$ GeV which corresponds to $2\pi/L$ on the N6
lattice. We keep the same value of $\vec{p}_\K$ on the other lattices by
introducing flavour-twisted boundary conditions \cite{tbc} for the strange
quark (cf.~Ref.~\cite{paper2}). 

Computing all-to-all propagators with a random source on every timeslice (``full
time dilution'') allows us to access all time separations in the two-point and
three-point functions. For more details on the measurements and analysis we
refer to the upcoming Ref.~\cite{paper1}.

\section{Bare form factor extraction by means of a combined fit}

Our goal is to extract the form factors by fitting to the three-point
correlation function, Eq.~(\ref{eq:c3}). We first determine the parameters of
the two-point correlation functions, Eqs.~(\ref{eq:c2ll}), (\ref{eq:c2hl}), and
use these as fixed input to estimate the form factors $\varphi_\mu$ from a
linear fit to \eq{eq:c3}. We then use all these parameters as initial values
to a ``global'' combined fit to Eqs.~(\ref{eq:c2ll})--(\ref{eq:c3}). We find
that in this way one obtains superior stability of the fit results with respect
to small changes of the initial values and fit ranges.

Our interest is mostly limited to the ground-state form factors
$\varphi_\mu^{(0,0)}$, however we find that for the safe extraction, free of
contamination by the excited states, we need to include more terms in the sums
of Eqs.~(\ref{eq:c2ll})--(\ref{eq:c3}). In these proceedings we keep only the
Kaon ground state but include two excited states for the $\Bs$ meson.

Clearly, a good choice for the initial values and a careful choice of fit ranges
is required to obtain stable combined fits. Let us briefly describe how this is
done in the following subsections.

\subsection{Two-point function fits}

\eq{eq:c2ll} for the two-point light-light correlator is taken in the limit of
infinite $T$. In practice, our lattices have finite $T$ and one has to take into
account the wrap-around state, giving
\begin{align}
\C^\K(t) \cong& (\kappa^{(0)})^2
(e^{-\EK{0}t}+e^{-\EK{0}(T-t)}),\label{eq:c2ll_nk1}
\end{align}
when $t$ and $T-t$ are large enough that we can neglect the contribution of the
excited states.

We select the time $\tfit{K2}{min}$ at which we start the fit by choosing the
smallest value of $t$ at which the (fitted) excited-state contribution is
smaller than 1/4 of the statistical uncertainty at that value of $t$. There is
no severe signal-to-noise problem in the Kaon sector, so we always use
$\tfit{K2}{max}=T/2$.

For the two-point heavy-light ($\Bs$) correlator, we have three different
smearings. Including the off-diagonal terms yields six independent correlators
in the symmetric $\C^{\Bs}_{ij}$ matrix. We first obtain the energies using
GEVP, with $t_0=\lceil t/2 \rceil$ (cf.~Refs.~\cite{gevp09,mb13}). Then we
determine the amplitudes in \eq{eq:c2hl} by first doing a linear fit to the
diagonal elements of $\C^{\Bs}$ to find the squares of the amplitudes and then
using these values as input to the non-linear fit for $\beta^{(n)}_i$ to all
elements of $\C^{\Bs}_{ij}$, including the off-diagonal ones.

\subsection{Safeguarding from finite-$T$ contributions in $\CBK$}

\begin{figure}[tpb]
\centering
\includegraphics[width=5.8cm]{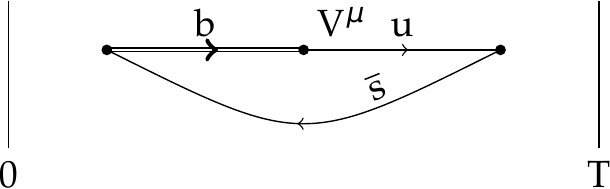}
\hspace{1.4cm}
\includegraphics[width=5.8cm]{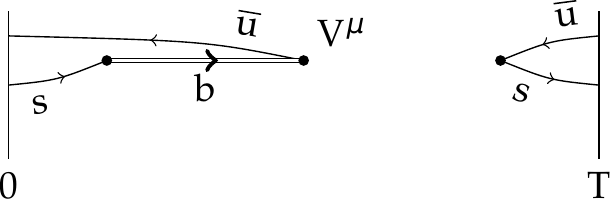}
\caption{Physical (left) and wrapper (right) contributions to $\CBK$.}
\label{fig:wrap}
\end{figure}

Due to the finite time extent of the lattice, we have to take into account the
``wrappers''' in the three-point functions -- at large enough times
$\CBK_{\mu,i}$ can be written as a sum of the ground-state contributions to
the two diagrams shown in Fig.~\ref{fig:wrap}:
\begin{equation}
\C^{\Bs\to\K}_{\mu,i}(t_\K,t_\Bs) \cong \kappa^{(0)}\varphi^{(0,0)}_\mu
\beta_i^{(0)} e^{-\EB{0}t_\Bs}e^{-\EK{0}t_\K}+
\kappa^{(0)}\xi_{\mu,i} e^{-E_{{\rm B}^*}t_\Bs} e^{-\EK{0}(T-t_\Bs-t_\K)},
\label{eq:wrap_full}
\end{equation}
where $\xi_{\mu,i}=\langle0|V_\mu|{\rm B}^*\rangle\langle{\rm
B}^*|P_{\rm hl}|\K\rangle$ is the unknown matrix element of the wrapper state
and $E_{{\rm B}^*}$ is the energy of the lightest heavy-light state contributing
to the wrapper diagram. In the static order we have $E_{{\rm B}^*}=\EB{0}$, but
at NLO it will be different.

To include the wrappers in the fit we need at least six extra parameters
$\xi_{\mu,i}$. Instead we choose to exclude these states by restricting the fit
region so that their contribution is negligible. To do that, for every given
value of $t_\Bs$, $\mu$, and $i$ we fit the three point function to the form
\begin{wrapfigure}{r}{7cm}
\includegraphics[width=7cm]{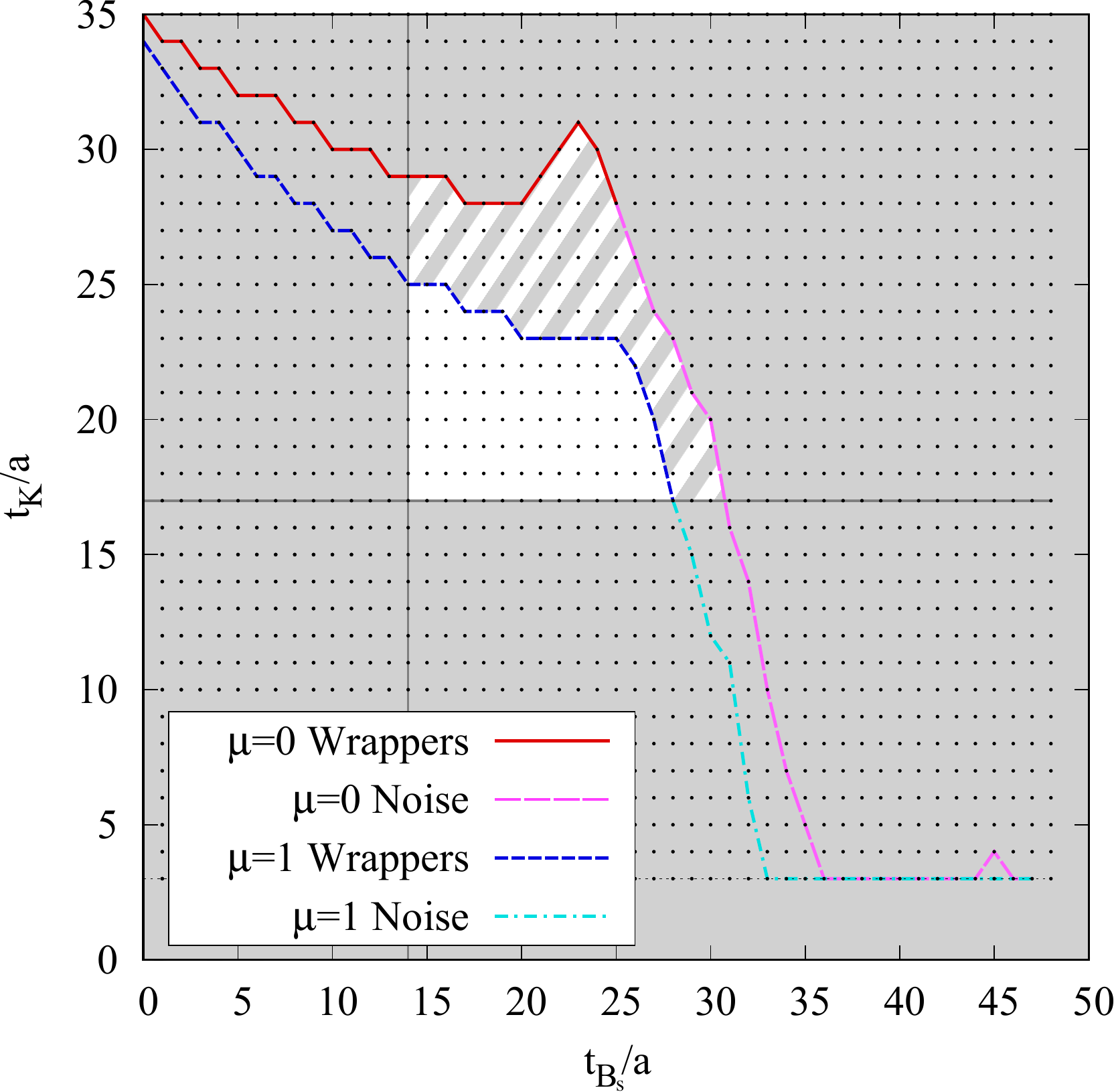}
\caption{The lines of $\tfit{K3}{max}$ for $\mu=0$ and $\mu=1$. The regions
disallowed for the fits (with $\tfit{K3}{min}$, $\tfit{B3}{min}$ from
\cite{paper2}) are in gray. The region allowed for $\mu=0$ but disallowed for
$\mu=1$ is in white-gray stripes.}
\label{fig:N6_noise_wrap}
\end{wrapfigure}
\begin{equation}
\C^{\Bs\to\K}_{\mu,i}(t_\K,t_\Bs) \cong
B_{\mu,i}e^{-E^\K_{\mu,i} t_\K}+C_{\mu,i}e^{+E^\K_{\mu,i} t_\K},
\end{equation}
with $B_{\mu,i}$ and $C_{\mu,i}$ being linear fit parameters (which one can
express in terms of the amplitudes and matrix elements of \eq{eq:wrap_full}) and
$E^\K_{\mu,i}$ being a non-linear fit parameter. Then we find
$\tfit{K3}{max,wr}$ as the last $t_\K$ for which the fitted wrapper contribution
to the function is smaller than 1/3 of its statistical uncertainty at that value
of $t_\K$. Final $\tfit{K3}{max}(\tb)$ can be chosen as the minimum of
$\tfit{K3}{max,wr}(\tb)$ and $\tfit{K3}{max,noise}(\tb)$, where the latter
excludes the points with a relative statistical error of $\CBK(\tk,\tb)$ larger
than 0.1.

The resulting $t^{\K3}_{\rm max}(t_\Bs)$ curves for the N6 ensemble are plotted
in Fig.~\ref{fig:N6_noise_wrap}. One clearly observes that the wrapper
contamination is more pronounced for $\mu=1$.



\subsection{Combined fit, stability}

Having determined the parameters in Eqs.~(\ref{eq:c2ll}), (\ref{eq:c2hl}) and
the estimates for the form factors $\varphi_\mu$ from linear fits to
Eq.~(\ref{eq:c3}), we use them as initial values for the combined non-linear
fit, i.e.\ we simultaneously fit the three equations for all values of $\mu$ and
all light-quark smearings for the $\Bs$ meson.

The temporal fit ranges are determined by suitable criteria described above,
except for three minimum times: $\tfit{B2}{min}$, $\tfit{B3}{min}$ and
$\tfit{K3}{min}$, which in our setup are independent of $\mu$ and the smearing
level. They are chosen such that the contributions from the excited states are
negligible.

\begin{figure}[tbp] \centering
\includegraphics[width=7.37cm]{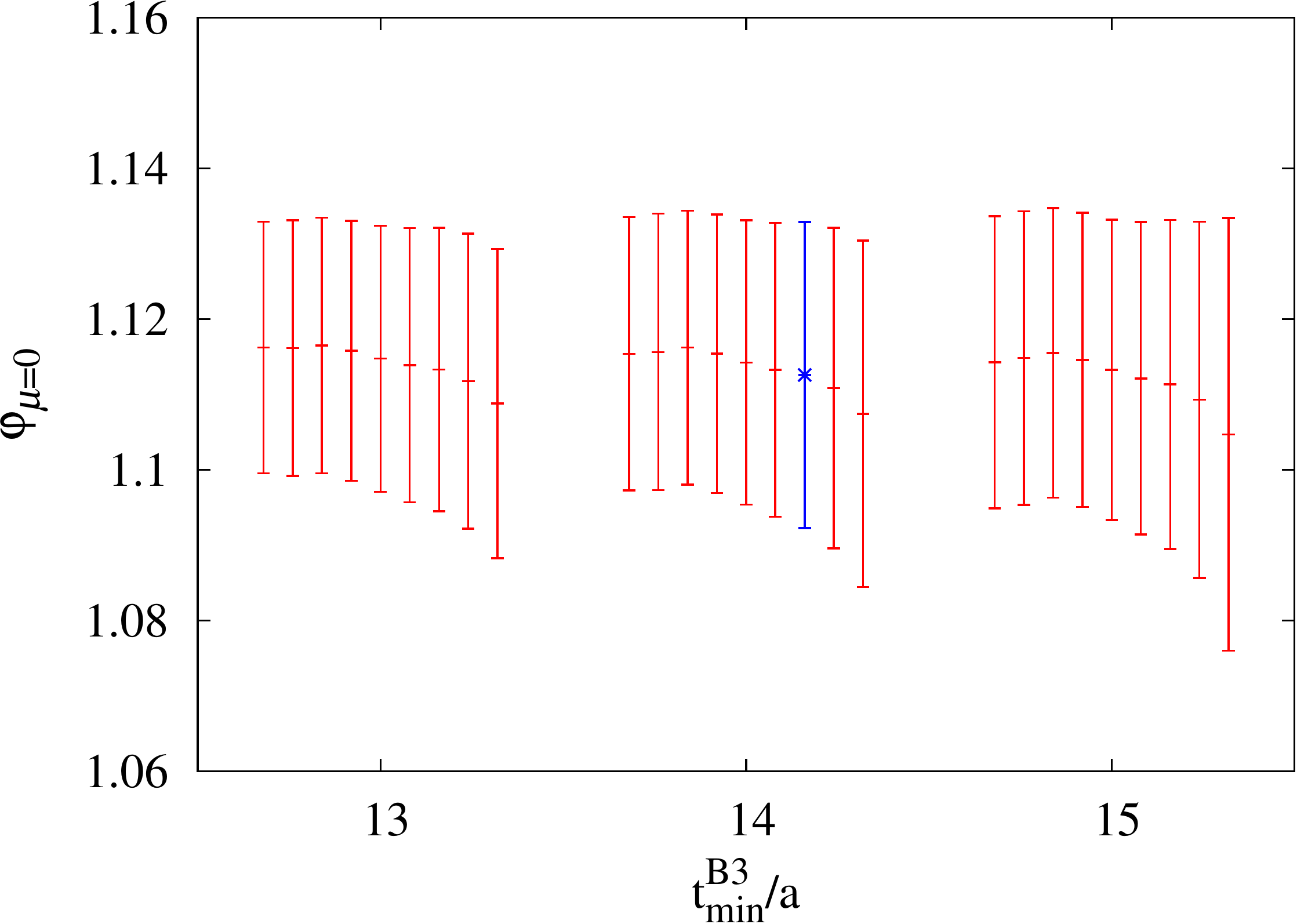}
\hspace{0.2cm}
\includegraphics[width=7.37cm]{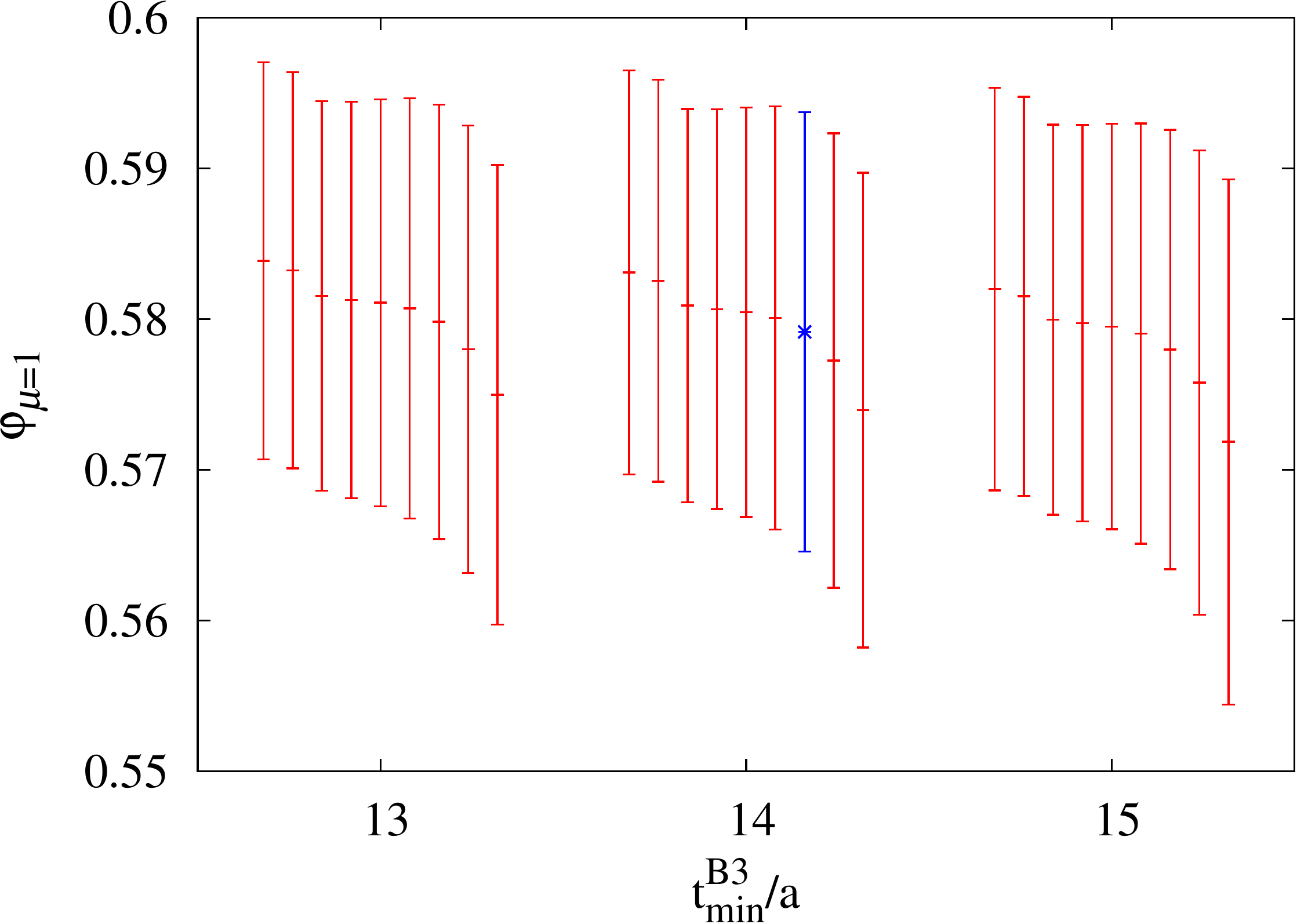} \caption{Stability of the
fit parameters $\varphi_0^{(0,0)}$ (left) and $\varphi_1^{(0,0)}$ (right) on
ensemble N6 (HYP2 discretization) with respect to variations of
$\tfit{B3}{min}/a$ (different groups) and of $\tfit{K3}{min} /a = 11\ldots19$
(within the groups). In the plot we fix $\tfit{B2}{min}=\tfit{B3}{min}-5a$. In
each panel, the value used to determine the bare form factor is marked with a
filled square.}
\label{fig:fit_stab}
\end{figure}

We check that the change of the fit results with respect to variations of the
fit ranges is negligible within the statistical errors, plotting the results on
the ``stability plots''. An example stability plot for the ground-state matrix
elements is presented in Fig.~\ref{fig:fit_stab}.

\section{Bare form factor extraction by the ratio method}

To cross-check the results obtained by the combined fit, we use the ratio
method. One can define many different ratios that converge to the desired form
factor in the limit of large $\tk,\tb$. Here, we consider three definitions
($\tau=\tk+\tb$):
\begin{align}
  \Ri_{\mu,i}(\tk,\tb) = &
    \frac{\C^{\Bs\to\K}_{\mu,i}(\tk,\tb)}{\big[\C^\K(\tau)
    \C^\Bs_{ii}(\tau)\big]^{1/2}} {\rm
    e}^{(\tilde{E}_\Bs-\tilde{E}_\K)\frac{\tb-\tk}{2}},\\[3pt]
    \Rii_{\mu,i}(\tk,\tb) = &
    \frac{\C^{\Bs\to\K}_{\mu,i}(\tk,\tb)}{\big[\C^\K(\tk)
    \C^\Bs_{ii}(\tb)\big]^{1/2}} {\rm
    e}^{\,\tilde{E}_\Bs\frac{\tb}{2}+\tilde{E}_\K\frac{\tk}{2}},\\[3pt]
    \Riii_{\mu,i}(\tk,\tb) = &
    \frac{\C^{\Bs\to\K}_{\mu,i}(\tk,\tb)}{\mathcal{N}^\K\C^\K(\tk)\,
    \mathcal{N}^\Bs_i\C^\Bs_{ii}(\tb)},
\end{align}
and restrict ourselves to the highest $\Bs$ smearing\footnote{We also analyzed
GEVP ratios, following Refs.~\cite{gevp09,gevp11}, but found no significant
improvement.}, and the case $\tk=\tb=t$.

Ratio $\Ri(t,t)$ has a particularly simple form and needs no extra
parameters -- it however comes at the price of working with $\C^\Bs(2t)$ in the
denominator, which results in more noisy behaviour at large $t$.

For $\Rii$ we find that one gets good statistical precision of the results when
using $\tilde{E}_\K=E^{(0)}_\K$, $\tilde{E}_\Bs=E^{(0)}_\Bs$ obtained from the
two-point function fits as described in the previous section. Also for $\Riii$
we use the fitted amplitudes $\mathcal{N}^\K = 1/\kappa^{(0)}$ and
$\mathcal{N}_i^\Bs = 1/\beta_i^{(0)}$.

\begin{figure}[tbp]
\centering
\includegraphics[width=7.37cm]{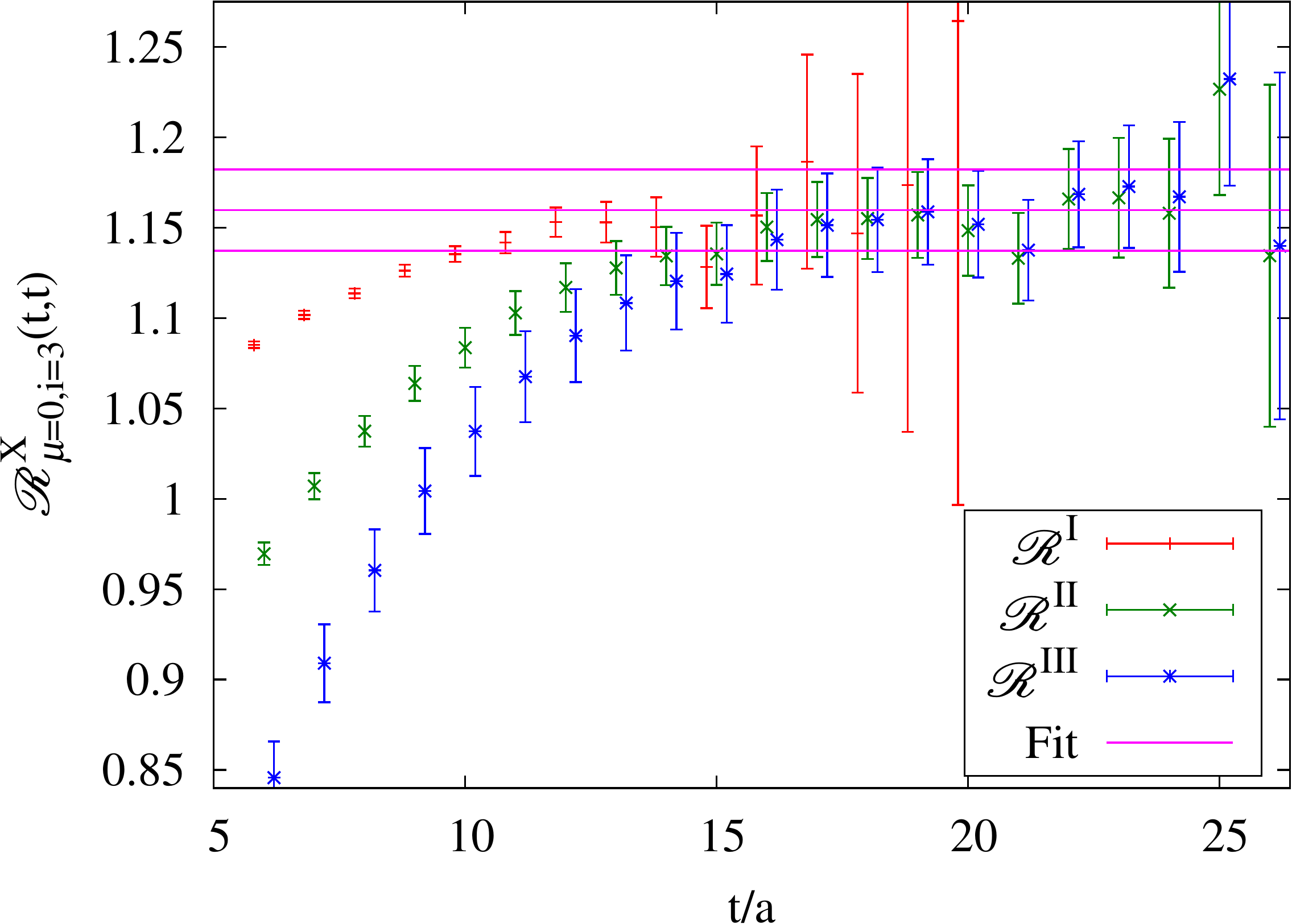}
\hspace{0.2cm}
\includegraphics[width=7.37cm]{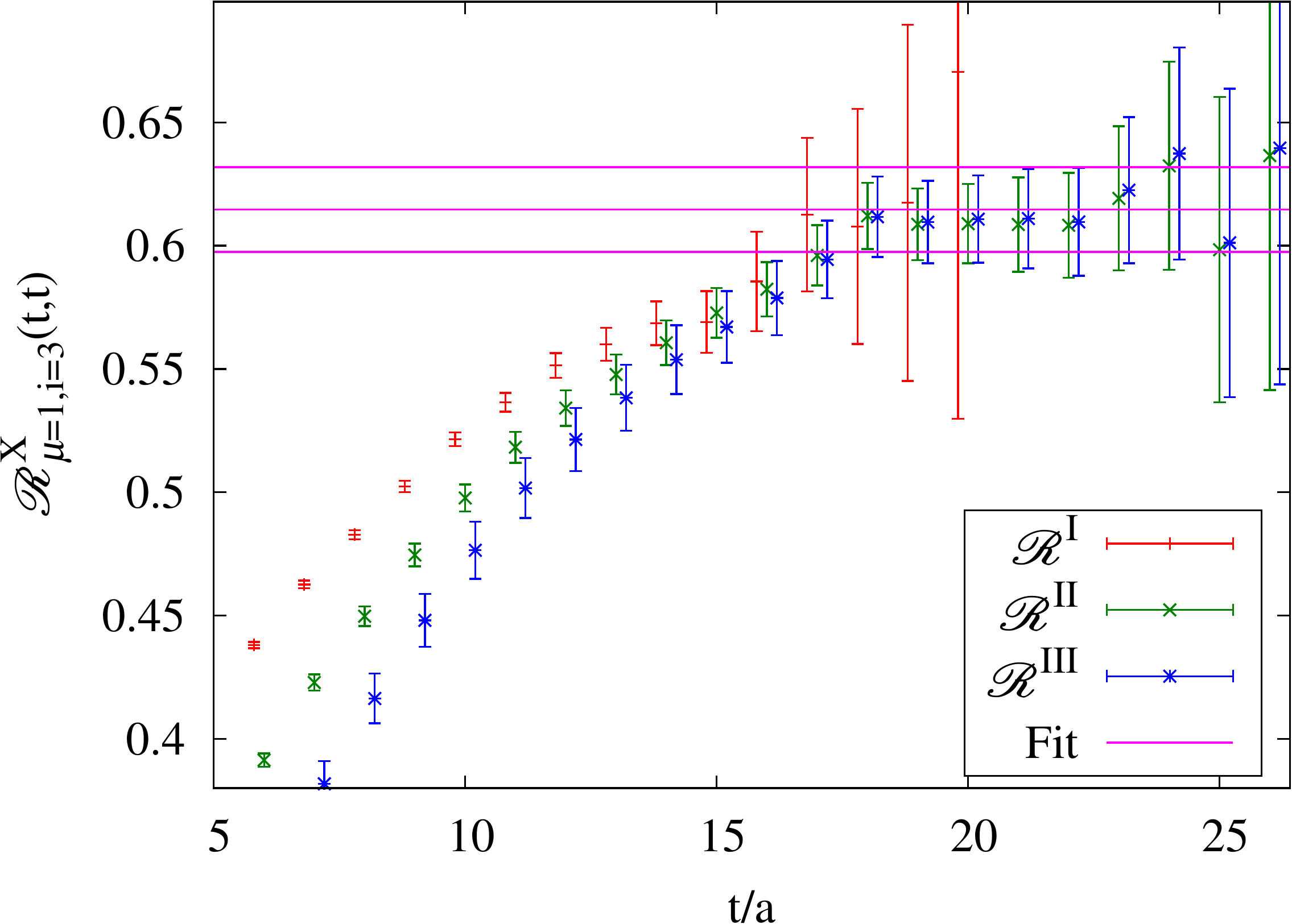}
\caption{Different ratios $\R^X_{\mu,i=3}(t,t)$ for the N6 ensemble (HYP1
discretization): $\mu=0$ (left) and $\mu=1$ (right), together with the combined
fit results. Results are slightly displaced on the horizontal axis for better
visibility.}
\label{fig:rcomp_tt}
\end{figure}

The results for the finest lattice spacing are presented in
Fig.~\ref{fig:rcomp_tt}. We see that $\Rii$ and $\Riii$ are nicely consistent
with the combined fit results in the vicinity of $0.8\,{\rm fm}\lesssim t
\lesssim 1\,{\rm fm}$. $\Ri_{\mu=0}$ has superior behaviour for small $t$ but
$\Ri_{\mu=1}$ does not and it becomes very noisy before reaching the plateau due
to the noise in $\C^\Bs(2t)$.

Improved convergence for $\Ri$ can be obtained by summing the ratio
\cite{sumrat}:
\begin{equation}
\mathcal{M}_{\mu,i}^{I}(\tau)=\partial_\tau\,
a\sum_{\tb}\Ri_{\mu,i}(\tau-\tb,\tb).
\vspace{-0.2cm}
\end{equation}
The asymptotic excited-state contaminations are then $\mathcal{O}(\tau\Delta
{\rm e}^{-\tau\Delta})$, where
$\Delta=\min(\EK{1}\!\!-\EK{0},\linebreak\EB{1}\!\!-\EB{0})$, as opposed to
$\mathcal{O}({\rm e}^{-\tau\Delta/2})$ in ordinary ratios \cite{gevp11}.

\begin{figure}[tbp]
\centering
\includegraphics[width=7.37cm]{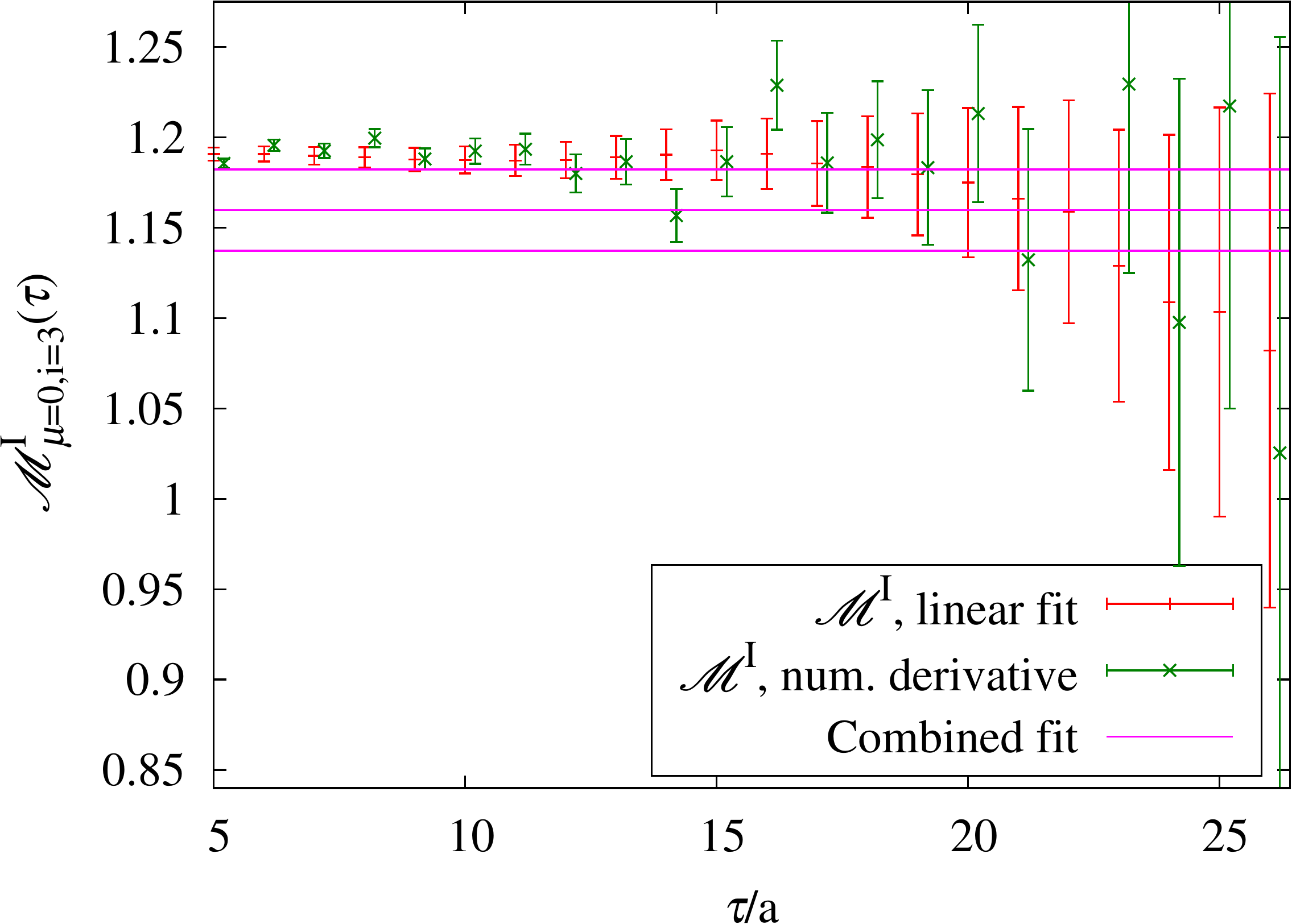}
\hspace{0.2cm}
\includegraphics[width=7.37cm]{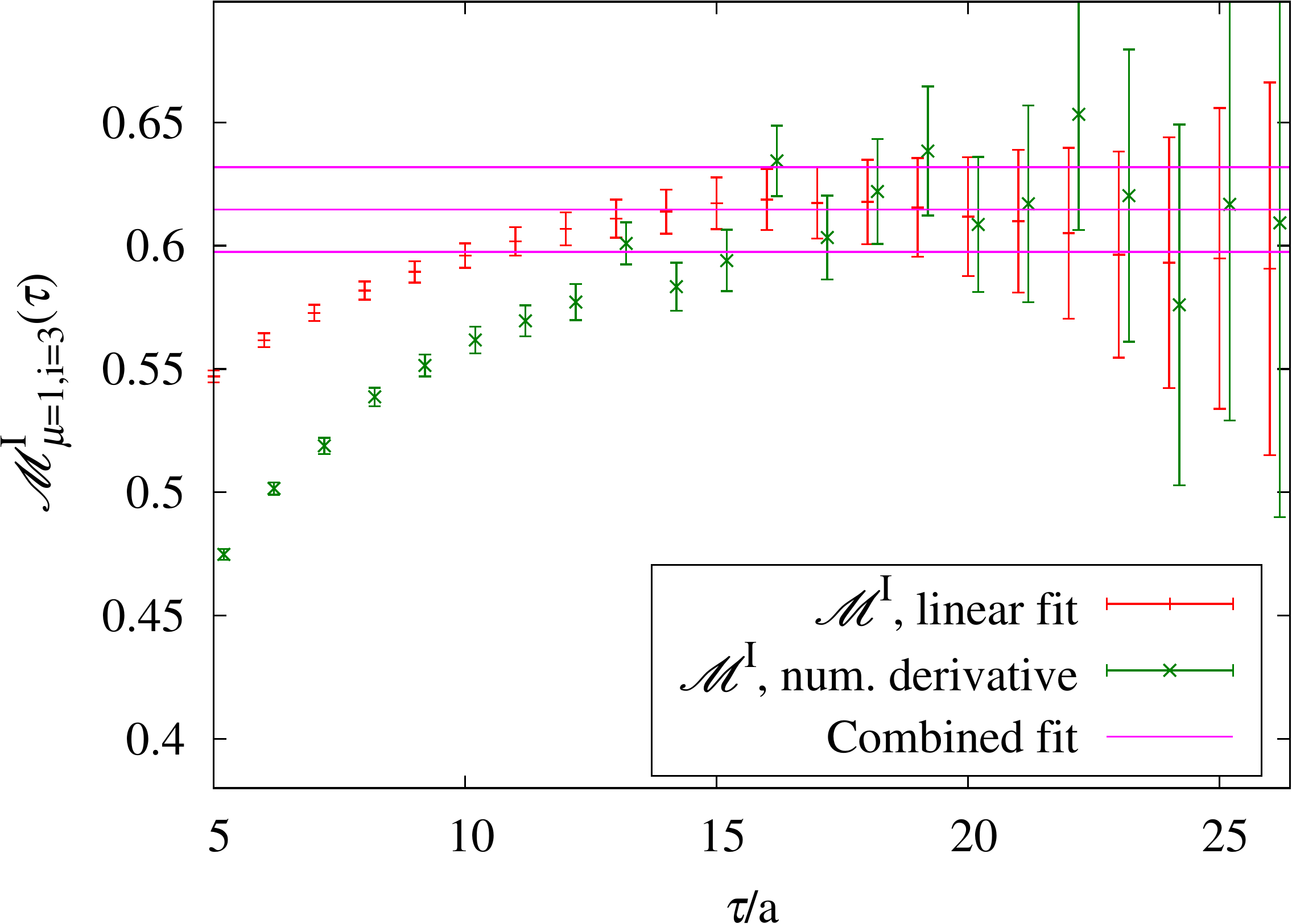}
\caption{Summed ratios $\mathcal{M}^I_{\mu,i=3}(t,t)$ for the N6 ensemble (HYP1
discretization): $\mu=0$ (left) and $\mu=1$ (right), obtained using two methods
described in text. The linear fit is done from $\tau$ to $\tau_{\rm max}=32a$.}
\label{fig:summed}
\end{figure}

In practice, the derivative can be calculated numerically or one can obtain
$\mathcal{M}_{\mu}^{I}$ from a linear fit to the sum (setting $\tau_{\rm max}$
small enough to avoid the influence of the wrappers).
Example results showing the two methods are presented in Fig.~\ref{fig:summed}.
We observe that in fact the convergence is improved.


\section{Conclusions and outlook}

We extract the bare matrix elements of the semi-leptonic $\Bs$ decay in the
static order of HQET. In this framework, renormalization
and matching to QCD can be performed non-perturbatively, such that the
continuum limit can be taken, as described in Refs.~\cite{deb16,paper2}.

Our numerical setup allows us to access all time separations in the two-point
and three-point correlation functions of interest, therefore giving us a very
good handle on the excited-state and finite-$T$ contributions.

To extract the matrix elements from the lattice data, we use two different
extraction methods: the combined fit and the ratio method. They give consistent
results which makes us confident in the robustness of the analysis.
The precision of our results for the bare matrix elements is approx. 2\%. The
final resulting precision of the static \textit{continuum-extrapolated} RGI
form factors is better than 5\% \cite{paper2}.

We estimate the systematic error from neglecting the subleading terms in
$1/m_{\rm b}$ to be of order 15\%. Therefore, the next step is to include the
$1/m_{\rm b}$ terms, which will reduce this systematic error to 1--2\%. The
required set of HQET parameters is being calculated in the parallel effort by
the ALPHA collaboration \cite{hei16}.

We also plan to include ensembles with a smaller pion mass, although we expect
the effects of the quark mass to be below our uncertainties in the case of
the $\Bs\to\K\ell\nu$ decay.


\begin{thebibliography}{99}
\bibitem{deb16} D.~Banerjee, ``\textit{Form factors in the $\Bs\to\K\ell\nu$
decays using HQET and the lattice}'', PoS(LATTICE2016) 292.
\bibitem{strategy} J.~Heitger and R.~Sommer, ``\textit{Nonperturbative heavy
quark effective theory}'', JHEP \textbf{0402} (2004) 022,
[arXiv:hep-lat/0310035].
\bibitem{hyp} M.~Della Morte, A.~Shindler, and R.~Sommer, ``\textit{On lattice
actions for static quarks}'', JHEP \textbf{0508} (2005) 051,
[arXiv:hep-lat/0506008].
\bibitem{gauss} S.~G\"usken et al.,
``\textit{Nonsinglet axial vector couplings of the baryon octet in lattice
QCD}'', Phys.~Lett.~\textbf{B227} (1989) 266.
\bibitem{mb13} F.~Bernardoni et al.,
``\textit{The b-quark mass from non-perturbative $N_f = 2$ Heavy Quark Effective
Theory at $O(1/m_h)$}'', Phys.~Lett.~\textbf{B730} (2014) 171,
[arXiv:1311.5498].
\bibitem{cls2} P.~Fritzsch et al., ``\textit{The strange quark mass and Lambda
parameter of two flavor QCD}'', Nucl.Phys. \textbf{B865} (2012) 397,
[arXiv:1205.5380].
\bibitem{paper2} F.~Bahr et al., ``\textit{Continuum limit of the leading-order
HQET form factor in $\Bs\to\K\ell\nu$ decays}'', Phys.~Lett.~\textbf{B757} (2016)
473, [arXiv:1601.04277].
\bibitem{tbc} P.~F.~Bedaque, ``\textit{Aharonov-Bohm effect and nucleon nucleon
phase shifts on the lattice}'', Phys.~Lett.~\textbf{B593} (2004) 82,
[arXiv:nucl-th/0402051].
\bibitem{paper1} F.~Bahr, D.~Banerjee, M.~Koren, H.~Simma, and R.~Sommer, in
preparation.
\bibitem{gevp09} B.~Blossier et al., ``\textit{On the generalized eigenvalue
method for energies and matrix elements in lattice field theory}'', JHEP
\textbf{0904} (2009) 094, [arXiv:0902.1265].
\bibitem{gevp11} J.~Bulava, M.~Donnellan, and R.~Sommer, ``\textit{On the
computation of hadron-to-hadron transition matrix elements in lattice QCD}'',
JHEP \textbf{01} (2012) 140, [arXiv:1108.3774].
\bibitem{sumrat} L.~Maiani et al.,
``\textit{Scalar densities and baryon mass differences in lattice QCD with
Wilson fermions}'', Nucl.~Phys.~\textbf{B293} (1987) 420.
\bibitem{hei16} M.~Della Morte et al., ``\textit{Non-perturbative matching of
HQET heavy-light axial and vector currents in $N_f = 2$ lattice QCD}'',
PoS(LATTICE2016) 199.
\end{thebibliography}
\end{document}